\documentclass[aps,showpacs,amsmath,showkeys,preprintnumbers,nofootinbib]{revtex4-2}
\usepackage{graphicx}
\usepackage{color}
\usepackage{amsmath}
\usepackage{amssymb}
\usepackage{multirow}

\makeatletter
\renewcommand\paragraph{\@startsection{paragraph}{4}{\z@}%
            {-2.5ex\@plus -1ex \@minus -.25ex}%
            {1.25ex \@plus .25ex}%
            {\normalfont\normalsize\bfseries}}
\makeatother
\def \be  {\begin{equation}}
\def \ee  {\end{equation}}
\def \ee  {\end{equation}}
\def \bea {\begin{eqnarray}}
\def \eea {\end{eqnarray}}

\newcommand{\nn}{\nonumber}

\begin{document}

\preprint{ECTP-2021-08}
\preprint{WLCAPP-2021-08}
\hspace{0.05cm}

\title{An appropriate statistical approach for nonequilibrium particle production}

\author{A. Tawfik}
\email{a.tawfik@fue.edu.eg}
\affiliation{Future University in Egypt (FUE), 5th Settlement, 11835 New Cairo, Egypt}

\author{E. R. Abou Elyazeed}
\affiliation{Physics Department, Faculty of Women for Arts, Science and Education, Ain Shams University, 11577 Cairo, Egypt}

\author{A. A. Alshehri}
\affiliation{University of Hafr Al Batin, Al Jamiah street, Hafar Al Batin 39524, KSA}

\author{H. Yassin}
\affiliation{Physics Department, College of Women for Arts, Science and Education, Ain Shams University, 11577 Cairo, Egypt}

\begin{abstract}

The incapability of thermal models to accurately reproduce the horn-like structure of the Kaon-to-pion ratio measured at AGS, SPS, and low RHIC energies, as well as confirmed in the beam energy scan program, has long been a persistent problem. This issue is believed to have arisen due to the inappropriate application of statistics, particularly the extensive additive Boltzmann-Gibbs (BG) statistics. The assumption that the analysis of particle production, a dynamic nonequilibrium process, should be primarily conducted using extensive BG or nonextensive Tsallis statistics, has proven to be an unsuccessful approach that has been followed for several decades.
By employing generic (non)extensive statistics, two equivalence classes $(c,d)$ emerge, thereby undermining the validity of any ad hoc assumption. Consequently, the degree of (non)extensivity exhibited by the statistical ensemble is determined by its own characteristics. This encompasses both extensive BG statistics, characterized by $(1,1)$, and nonextensive Tsallis statistics, characterized by $(0,q)$.
The energy dependence of light-, $\gamma_q$, and strange-quark occupation factor, $\gamma_s$, suggests that the produced particles are most appropriately described as a nonequilibrium ensemble. This is evidenced by a remarkable nonmonotonic behavior observed in the $\mathrm{K}^+/\pi^+$ horn, for instance. On the other hand, the resulting equivalence classes $(c,d)$ are associated with a generic nonextensivity related to extended exponential and Lambert-$W_0$ exponentially generating distribution function, which evidently arise from free, short- and long-range correlations.
The incorporation of generic nonextensive statistics into the hadron resonance gas model yields an impressive ability to rightfully reproduce the nonmonotonic $\mathrm{K}^+/\pi^+$ ratio.
\end{abstract}

\pacs{05.70.Ln, 05.65.+b, 25.75.-q, 05.90.+m} 
\keywords{Nonequilibrium and irreversible thermodynamics, Self-organized systems, Relativistic heavy-ion collisions, Other topics in statistical physics, thermodynamics, and nonlinear dynamical systems}

\date{\today}

\maketitle


\section{Introduction}
\label{intro}

Experiments conducted at AGS, SPS, and low RHIC energies have confirmed the presence of a nonmonotonic horn-like structure in the $\mathrm{K}^+/\pi^+$ ratio \cite{Gazdzicki:1998vd,E866:1999ktz,NA44:2002wha,NA49:2002pzu,NA49:2004iqm}. At energies below $\sqrt{s_{NN}}\lesssim 20~$GeV, the ratio exhibits a rapid increase followed by a rapid decrease, resulting in the appearance of a maximum. However, at energies above $\gtrsim 20~$GeV, such as those observed at RHIC and LHC, the $\mathrm{K}^+/\pi^+$ ratio appears to be nearly independent of energy, forming a plateau. The literature offers various interpretations for this phenomenon \cite{Gazdzicki:1998vd,Cleymans:2005zx,Andronic:2005yp,Tawfik:2005gk,Tomasik:2006qs,Chatterjee:2009km,Tawfik:2010uh,Nayak:2010uq}. However, the thermal models based on either extensive additive Boltzmann-Gibbs (BG) statistics or Tsallis-type nonextensive statistics \cite{Tawfik:2019alr} have not been successful in reproducing the nonmonotonic horn-like structure observed in the $\mathrm{K}^+/\pi^+$ ratio, as depicted in Fig. \ref{fig:0} and Fig. \ref{fig:2}. Despite this, there have been attempts to extend the thermal models in order to provide explanations for the underlying processes involved in particle production \cite{Shen:2019kil,Shen:2019zgi,Shen:2017pyo,Biro:2016dgd,Parvan:2010vn,Wilk:2014zka,Rybczynski:2012vj}.

To improve the capability of the thermal models; the ideal gas of identified hadronic resonances, one of the authors (AT) proposed i) a substantial change in the single-particle distribution function due to a dynamical change in the phase-space volume, especially at the phase transition, whose imprints seem to survive till the chemical freezeout, where the produced particles are detected \cite{Tawfik:2010uh}, ii) integrating nonequilibrium light- and strange-quark occupation factors, $\gamma_{q,s}$, in the phase-space volume \cite{Tawfik:2005gk,Tawfik:2016rrv,Tawfik:2017oyn},  iii) applying Tsallis-type nonextensive statistics \cite{Tawfik:2019alr}, and iv) deriving the ideal gas of hadron resonances away from equilibrium by assigning {\it eigen}-volume to each of its constituents \cite{Tawfik:2014eba,Tawfik:2013eua}. Castorina {\it et al.} assumed a suppression in the strange-quark occupation factor, $\gamma_s$, which accounts for reducing the production rates of the strange particles in $pp$, $pA$, $AA$ collisions at different energies, and found that the strangeness suppression disappeared with the onset of the color deconfinement and full-chemical nonequilibrium was subsequently attained \cite{Castorina:2017dvt,Castorina:2018zxi}. The occupation factors, $\gamma_{q,s}$, introduce controllers over the numbers of light and strange-quarks fitting within the phase-space element, that is, a similar role to that of the chemical potentials \cite{Yassin:2020zzy,Tawfik:2019gpc,Tawfik:2020gzw}. None of these ingratiates satisfactorily reproduced various particle ratios including $\mathrm{K}^+/\pi^+$ \cite{Tawfik:2018ahq,Tawfik:2016rrv,Tawfik:2017oyn}. 

Therefore, there is a conceptual reason for this. Imposing nonequilibrium (either as partial or full chemical nonequilibrium $\gamma_{q,s}$ or as finite sizes of the constituents of the ideal gas) in an equilibrium ensemble of BG-type extensive additive statistics seems to be unsuccessful. Even if Tsallis-type nonextensivity is assumed, the matching of nonequilibrium with nonextensivity would improve. However, the price to be paid in return is large. This is not less than losing the meaning of thermodynamic quantities, such as the freezeout temperature \cite{Deppman:2012qt,Biro:2016myk,NasserTawfik:2016sqs}. In this regard, it should be highlighted that the Tsallis-type nonextensivity is limited to a one-dimensional line within the entire space of (non)extensivity utilized in the present calculations, in which both the BG and Tsallis statistics are merely special cases. 

Relativistic high-energy collisions are assumed to undergo several {\it critical} processes, such as deconfinement and hadronization. A radical change in the properties, symmetries, and degrees of freedom (dof) of the QCD matter depends on the collision energy. The horn-like structure of the kaon-to-pion ratio, which normalizes  strange to light quark flavors, likely signals the onset of nonextensive particle production and nonequilibrium phase transition that might dynamically enhance the strange quark flavor or suppress the light quark flavor, or vice versa. The present script suggests that the incapability of the {\it extensive additive} thermal models in reproducing the horn-like structure measured at AGS, SPS, and low RHIC energies is due to inappropriate statistics. Generic (non)extensive statistics, autonomously describes the equivalence classes of {\it generalized} entropies in the entire $(c,d)$-plane \cite{Thurner1}. At chemical freezeout, where detection of the produced particles occurs, the statistical ensemble is dynamic and probably violates the fourth Shannon-Khinchin axiom \cite{NasserTawfik:2016sqs}. Even if this is not the case and the produced particles are in full-chemical nonequilibrium, generic (non)extensive statistics are also applicable. Extensive or non-extensive ensemble and equilibrium or out-of-equilibrium systems are properly encoded by the equivalence classes $(c,d)$ including extensive BG and Tsallis-type non-extensive statistics, which are characterized by $(c,d)=(1,1)$ and $(c,d)=(0,q)$, respectively \cite{Tawfik:2017bul}. 

The remainder of this paper is organized as follows. A generic (non-)extensive approach is reviewed in Section {\ref{sec:app}}. The energy-dependence of the equivalence classes $(c,d)$, the light- and strange-quark occupation factors $\gamma_{s,q}$, and the $\mathrm{K}^+/\pi^+$ ratio is presented in section \ref{sec:cd}, \ref{sec:gamm}, and \ref{sec:kpi}, respectively. Section \ref{sec:Diss_concl} presents the conclusions. Appendices \ref{sec:expf}, \ref{sec:largemasssgma} and \ref{sec:cdcountors} elaborate on the probability distribution, entropy, exponential and logarithmic functions in Boltzmann, Tsallis, and generic (non)extensive statistics, the impacts of the inclusion of very high-mass resonances and missing states in the hadron resonance gas model, and the contour plots of the quality of fit for $\gamma_{q,s}$. The quality of the fit for equivalence classes $(c,d)$ is shown in Fig. \ref{fig:1b}.


\section{Generic (non)extensive statistical approach}
\label{sec:app}

In the domain of generic (non)extensive statistics, the partition function is formulated as \cite{Tawfik:2018ahq,Tawfik:2017bsy}
\bea
\ln\, Z(T,\mu) &=& \pm V\, \sum_i\, \frac{g_i}{(2\, \pi)^3}\, \int\, \ln\left[1\pm\varepsilon_{c,d,r}(x_i)\right]\; d^3\, {\bf p}, \label{eq1}
\eea
where $x_i=[\mu_{i}-E_i({\bf p})]/T$. The quantity $E_i(p)=\sqrt{{\bf p}^2+m_i^2}$ represents the dispersion relation of the $i$-th particle. Here, $\mu_i$ stands for the chemical potential of that particle, which can be expressed as $\mu_i=B_i \mu_B + S_i\mu_S+ ...$, where $S_i$ and $B_i$ are the corresponding strangeness and baryon quantum numbers, and $\mu_S$ and $\mu_B$ are the strangeness and baryon chemical potentials, respectively. In this context, $g$ represents the degeneracy factor, $V$ denotes the volume of the fireball, ${\bf p}$ symbolizes the momentum, and $\pm$ indicates fermions and bosons, respectively. The extended exponential function $\varepsilon_{c,d,r}(x_i)$ is defined as \cite{Thurner1,Thurner2}
\bea
\varepsilon_{c,d,r}(x_i)=\exp\left[ \frac{-d}{1-c} \left(W_k\left[B\left(1-\frac{x_i}{r}\right)^{\frac{1}{d}}\right]-W_k[B]\right)\right], \label{eq:epsln}
\eea
where $W_k$ is the Lambert-$W_k$ function which has real solutions, at $k=0$ with $d\geq 0$, and at $k=1$ with $d<0$.
\be
B=\frac{(1-c)r}{1-(1-c)r} \exp\left[\frac{(1-c)r}{1-(1-c)r}\right],
\ee
where $r=(1-c+cd)^{-1}$ with $c$, $d$ are two critical exponents defining the equivalence classes for both types of extensive and nonoextensive statistical ensembles. The latter are statistical systems violating the fourth Shannon-Khinchin axiom; the generalized Shannon additivity \footnote{If $p_{ij} \geq 0$, $p_i=\sum_{j=1}^{m_i} p_{ij}$ and $\sum_{i=1}^{n} p_{i}=1$, where $i=1,\cdots,n$ and $j=1,\cdots,m_i$, then $S(p_{11},\cdots,p_{nm_n}) = S_1(p_{1},\cdots,p_{n}) + \sum_{i=1}^{n} p_i S_1\left(\frac{p_{i1}}{p_i}\cdots,\frac{p_{i m_i}}{p_i}\right)$. Accordingly, $S_n$ is the Shannon entropy given by $S_n(P)=\tau \cdot \sum_{k=1}^{n} p_k \log_2 p_{k}$ with $\tau<0$, that is, the entropy of a system, which can be divided into subsystems $A$ and $B$ is given as $S_A$ and the expectation value of $S_B$ but conditional on $A$; $S_{nm}(AB)=S_n(A)+S_m(B|A)$, where $S_m(B|A)=\sum_k p_k \cdot S_m(B_k)$ \cite{Tawfik:2017bsy}. The generalized entropy reads $S_{c,d}[p] \propto \sum_{i=1}^{\Omega} \Gamma(d+1,1-c \log(p_i))$, where the constants $(c,d)$ characterize the universality class of entropy of the system completely in the thermodynamic limit and specify its distribution functions. The fourth Shannon-Khinchin axiom corresponds to Markovian processes, which are random processes whose future is independent of the past, and the natural stochastic analogy of the deterministic processes described by differential equations.}, particularly in their large size limit, in which two asymptotic properties of the generalized entropies are associated with one scaling function each. Each scaling function is characterized by one exponent defining the equivalence relations of the entropic forms, that is, two entropic forms are equivalent if their exponents are the same.

The equivalence classes $(c,d)$ which are two exponents that determine two scaling functions with two asymptotic properties, are intrinsic properties of all types of underlying statistics. For instance, extensive BG and Tsallis-type non-extensive statistics assume $(c,d)=(1,1)$ and $(0,q)$, respectively, \cite{Thurner2}. The remaining $c$–$d$ space is inhabited by various types of nonextensivity or superstatistics, enabling us to reveal nonequilibrium phenomena such as critical endpoint, correlations, fluctuations, and higher-order cumulants. The present script suggests that the horn of the kaon-to-pion ratio likely signals the onset of non-extensive particle production and the impact of non-equilibrium phase transition \cite{Tawfik:2018sji}. The inability of the {\it equilibrium} thermal models to reproduce the enhancement in both the particle yields, kaons, and pions, Fig. \ref{fig:0} and Fig. \ref{fig:2}, could be interpretted due to inappropriate statistics \cite{Tawfik:2014eba,Tawfik:2010uh,Tawfik:2017bsy,NasserTawfik:2016sqs,Tawfik:2018ahq}, i.e., either extensive BG or Tsallis-type nonextensive statistics. From Eq. (\ref{eq1}), both the Tsallis and Boltzmann-Gibbs statistics are clear. Deriving  $\ln Z$ of Tsallis $(0,q)$ and Boltzmann-Gibbs $(1,1)$ statistics is an unambitious exercise.

The generic partition function, which includes the occupation factors of light- and strange-quarks denoted as $\gamma_q$ and $\gamma_s$ respectively, can be expressed as
\bea
\ln\, Z(T,\mu) &=& \pm V\, \sum_i\, \frac{g_i}{(2\, \pi)^3}\, \int\, \ln\left[1\pm(\gamma_q^{n_q})_i\;(\gamma_s^{n_s})_i\;\varepsilon_{c,d,r}(x_i)\right]\; d^3\, {\bf p}, \label{eq2}
\eea
where $n_q$ and $n_s$ are the number of light and strange quarks, respectively \footnote{Given our focus on the QCD phase diagram, it is possible to disregard the impact of the heavier quarks in our analysis.}. As BG statistics assume extensivity, additivity, and equilibrium particle production, $\gamma_q$ and $\gamma_s$ can plainly be seen as additional variables imposed on BG statistics. However, they are essential ingredients for (non)extensive statistics. \footnote{This should not necessarily be limited to the Tsallis-type. The entire $c$–$d$ space characterizes various types of superstatistics, including extensive and non-extensive ensembles.}. In generic (non)extensive statistics, the degree of non-extensivity is subjectively determined by the changeable statistical nature of the system of interest at various energies. 

In this context, it is important to note that the statistical characteristics of a group of individual entities are determined by a distribution function that describes the likelihood of finding an entity within a small region of phase-space. This probability is directly related to both the volume of the phase-space element and the distribution function itself. The distribution function is expected to satisfy the Boltzmann-Vlasov (BV) master equation, which represents the semiclassical approximation of a time-dependent Hartree-Fock theory achieved through the Wigner transform of the one-body density matrix. This equation can be utilized to analyze the behavior of the components within the system. At higher energy levels, the phase-space volume and its configurations undergo noticeable alterations, leading to changes in the single-particle distribution function \cite{Tawfik:2010uh}. 

The occupation factors $\gamma_{q,s}$ serve as regulators for the number of quarks that can occupy the phase-space element. They can be compared to the chemical potentials $a=\ln E - \ln T - \ln N$, where $\lambda=\exp(-a)$ \cite{Tawfik:2010aq}. When $\gamma_s=\gamma_q=1$, the statistical ensemble represents a system in both chemical and thermal equilibrium. However, if either $\gamma_s$ or $\gamma_q$ deviates from unity while the other remains at equilibrium, the statistical system is transformed into a state of partial-chemical nonequilibrium. In the case of $\gamma_s\neq1$ and $\gamma_q\neq1$, the statistical ensemble is evidently in an out-of-equilibrium state.

When considering the thermodynamic limit, one can calculate the number density by examining the partition function, as indicated in Eq. (\ref{eq2}) \cite{RAFELSKI,Tawfik:2018ahq}, 
\bea
\textit{n}(T,\mu) &=& \pm \sum_{i}\, \frac{g_i}{2 \pi^2} \int_0^\infty \frac{(1-c)^{-1}\, (\gamma_q^{n_q})_i\;(\gamma_s^{n_s})_i\;\varepsilon_{c,d,r}(x_i) \; W_0\left[B(1-\frac{x_i}{r})^{\frac{1}{d}}\right]}{\left[1\pm(\gamma_q^{n_q})_i\;(\gamma_s^{n_s})_i\;\varepsilon_{c,d,r}(x_i)\right] \left(r-x_i\right) \left(1+W_0\left[B(1-\frac{x_i}{r})^{\frac{1}{d}}\right]\right)}\; \textbf{p}^2 d\textbf{p}.   \hspace{5mm}
\label{FDn1}
\eea 
At the chemical freezeout, which occurs throughout the QCD phase diagram under both extensive BG and generic (non)extensive statistics, hadron resonances decay into stable particles, $n_i^{direct}$, or other resonances $n_j$. This decay process impacts the determination of specific particle yields, such as pion and Kaon
\bea
\langle n_i^{final}\rangle &=& \langle n_i^{direct}\rangle + \sum_{j\neq i} b_{j\rightarrow i} \langle n_j\rangle. \label{FDn2}
\eea
The branching ratio $b_{j\rightarrow i}$ represents the probability of the decay of the $j$-th resonance into the $i$-th particle.

In this study, we initially consider partial-chemical nonequilibrium, where $\gamma_q$ is assumed to be equal to 1. On the other hand, the value of $\gamma_s$ is determined through statistical fits of the generic (non)extensive calculations of the $\mathrm{K}^+/\pi^+$ ratio. These calculations are performed over a wide range of center-of-mass energies $\sqrt{s_{NN}}$ and are compared to experimental measurements, as shown in Figure \ref{fig:1} panel (a). 

Furthermore, we also investigate full-chemical nonequilibrium, where both $\gamma_q$ and $\gamma_s$ are not equal to 1. Again, the values of $\gamma_q$ and $\gamma_s$ are determined through statistical fits of the $\mathrm{K}^+/\pi^+$ ratio. The results of these fits are presented in Figure \ref{fig:1} panels (b) and (c). 

The degree of (non)extensivity is autonomously determined by the statistical ensemble used in the analysis. The findings obtained from these calculations will be thoroughly discussed in the subsequent section of this work.

\section{Results}
\label{sec:res}

Through the use of statistical fits involving various particle ratios, the parameters $T$ and $\mu_B$ at the chemical freezeout can be determined. This freezeout is characterized by the conservation of quantum numbers such as baryon, strangeness, and electric charge. The values of $T$ and $\mu_B$ are expressed as functions of collision energies and are thermodynamically conditioned, with a constant entropy density of $s/T^3=7$ as an example. These fits can be performed using either extensive BG or generic (non)extensive statistics. 

In addition to determining $T$ and $\mu_B$, the energy dependence of equivalence classes $(c,d)$ and the occupation factors $\gamma_q$ and $\gamma_s$ for light and strange quarks have also been investigated using generic (non)extensive statistics. These investigations have been carried out at various center-of-mass energies $\sqrt{s_{NN}}$. 

The three sets of parameters, namely the freezeout parameters, the equivalence classes, and the occupation factors, are then used as inputs in calculations for the $\mathrm{K}^+/\pi^+$ ratio. At a given $\sqrt{s_{NN}}$, which is related to $\mu_B$ and the freezeout temperature $T$, the corresponding equivalence classes and occupation factors are substituted into the relevant equations. These equations, specifically Eq. (\ref{FDn1}) and Eq. (\ref{FDn2}), are used to determine the number density of a specific particle.

\subsection{Equivalence classes ($c,d$) and nonextensivity}
\label{sec:cd}

\begin{figure}[htb]
\center
\includegraphics[width=0.5\textwidth]{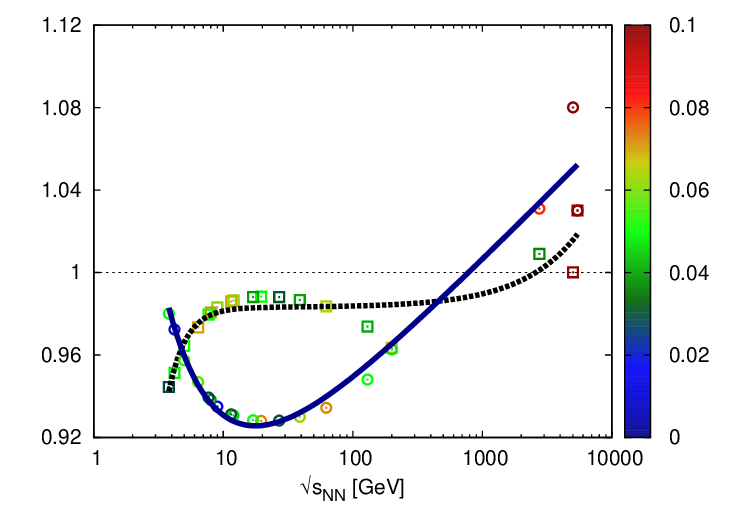}
\caption{The contour plot exhibits the relationship between the equivalence classes $(c,d)$ and $\sqrt{s_{NN}}$. In this plot, the dashed curve represents $c$ according to Eq. (\ref{eq:c}), while the solid curve corresponds to $d$ based on Eq. (\ref{eq:d}). A horizontal line is used to denote extensive BG statistics, with a thin dashed line indicating the specific equivalence classes of extensive BG statistics, namely $(c,d)=(1,1)$. The colored band on the left side of the plot illustrates the statistical errors.
}
\label{fig:1b}
\end{figure}

In Fig. \ref{fig:1b}, the squares and circles represent the resulting equivalence classes $(c,d)$ at different energies \cite{Tawfik:2018ahq}. The parameterization of these results can be accurately described by the following expressions:
\begin{eqnarray}
c(\sqrt{s_{NN}}) &=& \left[a_1 + \left(\sqrt{s_{NN}}\right)^{-b_1}\right]^{-c_1}\; d_1 \sqrt{s_{NN}}, \label{eq:c}\\ 	
d(\sqrt{s_{NN}}) &=& \left[a_2 - \left(\sqrt{s_{NN}}\right)^{-b_2}\right]^{c_2}\; \exp\left(\frac{d_2}{\sqrt{s_{NN}}}\right), \label{eq:d}
\end{eqnarray}
where $a_1= 1.005\pm0.01$, $b_1=3.242\pm0.106$, 
$c_1=3.247\pm0.24$, $d_1=6.526 \times 10^{-06}\pm1.003 \times 10^{-06}$, $a_2=1.947\pm0.0006$, $b_2=0.008\pm0.00015$, $c_2=3.493\pm0.0739$, and $d_2=0.506\pm0.019$. Both expressions are respectively depicted as dashed and solid curves in Fig. \ref{fig:1b}.

At AGS energies, the parameter $d$ initially starts at a high value close to unity and then rapidly decreases, while the parameter $c$ begins at approximately 0.94 and shows an exponential growth as the energy increases. When the center-of-mass energy $\sqrt{s_{NN}}$ is around 4±1 GeV, both $c$ and $d$ reach a similar value of approximately 0.97. Moving to SPS and lower RHIC energies, $c$ reaches its maximum value at around $\sqrt{s_{NN}}$ = 10±1 GeV, while $d$ continues to decrease and approaches its minimum value of around 0.93±0.02 at $\sqrt{s_{NN}}$ = 20±2 GeV. At higher RHIC and LHC energies, $c$ stabilizes at a plateau around 0.985. For $\sqrt{s_{NN}}$ greater than 20 GeV, $d$ transitions to an exponential increase with energy. Similarly, $c$ also exhibits an exponential increase at LHC energies, causing both parameters to surpass unity, thereby violating the second Shannon-Khinchin axiom.

The level of (non)extensivity can be ascertained through the equivalence classes $(c,d)$, which were most accurately described by the energy-dependent Eqs. (\ref{eq:c}) and (\ref{eq:d}), and illustrated in Fig. \ref{fig:1b}. It is observed that the resulting equivalence classes are predominantly positive and near to unity. This indicates that:
\begin{enumerate}
\item the underlying statistical framework is limited in scope at BG, particularly at AGS and SPS energy levels, and does not exhibit Tsallis-type nonextensive statistics, 
\item at AGS, SPS and RHIC energies, the statistical systems holds first \footnote{The Shannon entropy function $S_n(p)=\tau \cdot \sum_{k=1}^{n} p_k \log_2 p_{k}$, with $\tau<0$, exhibits continuity within $\Delta_n$, indicating that the entropy is a continuous function of $p$.}, second \footnote{The Shannon entropy is maximized for the uniform distribution $U_n=(1/n,\cdots,1/n)\in \Delta_n$, indicating that $S_n(p)\leq S_n(U_n)$ for every $p\in \Delta_n$ \cite{Tawfik:2017bsy}.} and third \footnote{The Shannon entropy is expandable; $S_{n+1}(p_1,p_2,\cdots,p_n,0)=S_n(p_1,p_2,\cdots,p_n)$ for any $(p_1,p_2,\cdots,p_n)\in \Delta_n$. This means that incorporating a state with zero probability into a system will not alter the entropy.} Shannon-Khinchin axioms but violates the fourth Shannon-Khinchin axiom, 
\item at LHC energies, both second and fourth Shannon-Khinchin axioms are violated, and d) the corresponding entropy is Lambert-$W_0$ exponential; exponentially generating function $W_0(x)=\sum_{n=1}^{\infty} (-n)^{n-1} x^n/ n!$.
\end{enumerate}

Even a small deviation from unity as shown in Figure \ref{fig:1b}, the values of $(c,d)$ play a crucial role in deriving the entire statistical system based on extensivity, additivity, and equilibrium. The Boltzmann statistics is specifically associated with $(c,d)=(1,1)$. Any departure from $(c,d)=(1,1)$ leads to the replacement of Boltzmann distribution, entropy, exponential, and logarithm functions with their counterparts that characterize generic (non)extensive statistics, as detailed in Appendix \ref{sec:expf}. The extended exponential and Lambert-$W_o$ exponentially generating distribution function exhibit significant distinctions from the standard exponential and distribution functions.

\subsection{Nonequilibrium quark occupation factors}
\label{sec:gamm}

\begin{figure}[htb]
\center
\includegraphics[width=7cm]{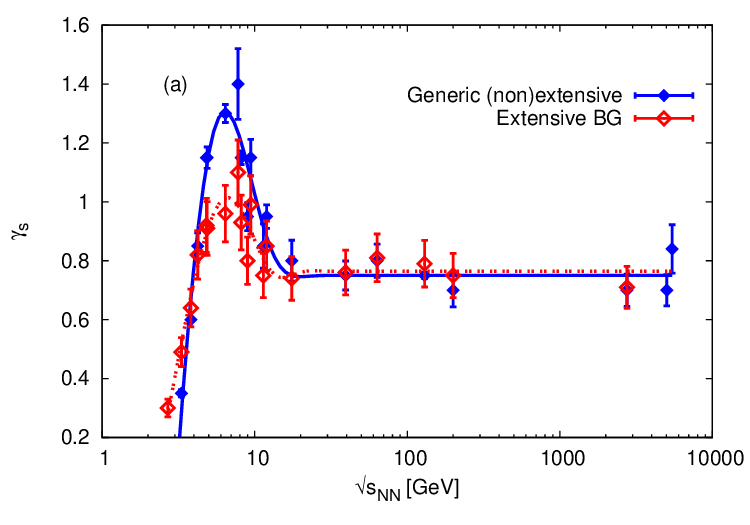} \\
\includegraphics[width=7cm]{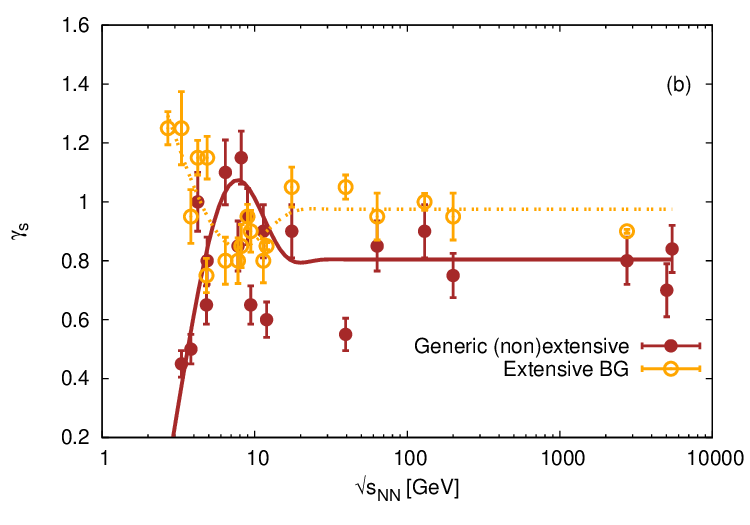}
\includegraphics[width=7cm]{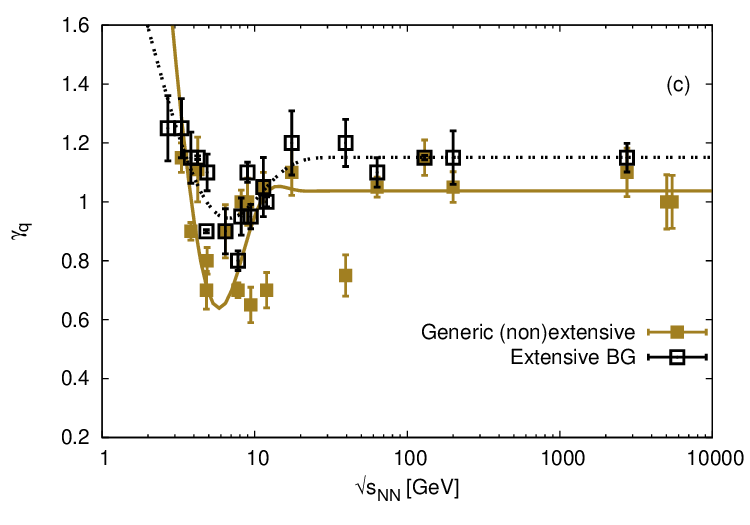}
\caption{In the top panel, $\gamma_s$ is expressed as a function of $\sqrt{s_{NN}}$, with $\gamma_q$ set to 1. On the other hand, in the bottom panel, both $\gamma_s$ (left) and $\gamma_s$ (right) are considered as fitting variables. Our results from the statistical fits of the measured $\mathrm{K}^+/\pi^+$ using generic (non)extensive calculations are represented by solid symbols with errors, while the statistical fits to the extensive BG statistics are shown by open symbols with errors. The curves corresponding to Eq. (\ref{eq:ourFit}) from our study and references \cite{Tawfik:2016rrv,Tawfik:2017oyn} are depicted by solid and double-dotted lines, respectively.
}
\label{fig:1} 
\end{figure}

In the context of partial- and full-chemical nonequilibrium, the ratios of different particles are measured at various collision energies ($\sqrt{s_{NN}}$) and then compared to the calculations based on generic (non)extensive statistics. At a specific collision energy, the corresponding equivalence classes $(c,d)$, as discussed in section \ref{sec:cd}, along with the freezeout parameters $\mu_B$ and $T$, are incorporated into the expression for the number density of a particular particle yield. By adjusting the fitting parameters, namely $\gamma_s$ for partial-chemical nonequilibrium with $\gamma_q$ fixed at 1, or both $\gamma_s$ and $\gamma_q$ for full-chemical nonequilibrium, the calculated ratios of various particles in generic (non)extensive statistics are matched to the experimentally measured ratios. The overall energy dependence of both $\gamma_q$ and $\gamma_s$ is illustrated in Fig. \ref{fig:1}
\bea
\gamma_{s,q}(\sqrt{s_{NN}}) &=& \alpha \, \exp\left(-\beta \, \sqrt{s_{NN}}\right)\, \sin\left(\nu \, \sqrt{s_{NN}} + \eta \right) + \zeta. \label{eq:ourFit} 
\eea
The parameters $\alpha,\;\beta,\;\nu,\;\eta,\;\zeta$ play a crucial role in distinguishing between partial- ($\gamma_q=1$ while $\gamma_s$ is the fitting variable) and full-chemical nonequilibrium, as shown in Tab. \ref{Tab1}.
 
\begin{table*}[htb]
\small
\begin{center}
\begin{tabular}{|c|c|c|c|c|c|c|}  \hline
\multicolumn{1}{ |c| }{\multirow{1}{*}{nonequilibrium}} &
   & $\alpha$ & $\beta$ & $\nu$ & $\eta$ & $\zeta$ \\ \hline
    \multirow{1}{*}{partial at $\gamma_q=1$} & $\gamma_s$ & $10.373\pm 0.625$ & $0.361\pm 0.009$ & $0.233\pm0.003$ & $5.362\pm 0.014$ & $0.751\pm 0.018$ \\ \hline
 \multirow{2}{*}{full} & $\gamma_s$ & $2.913\pm 0.259$ & $0.262\pm 0.062$ & $-0.257\pm 0.018$ & $4.353\pm 0.081$ & $0.804\pm 0.025$ \\
     \cline{2-7} & $\gamma_q$ & $-5.264\pm 0.256$ & $0.381\pm 0.018$ & $0.373\pm 0.017$ & $4.88\pm 0.071$ & $1.037\pm 0.041$  \\ \hline
\end{tabular}
\caption{Adjusting the parameters that govern the relationship between $\gamma_{s,q}$ and $\sqrt{s_{NN}}$ can be achieved by utilizing Equation (\ref{eq:ourFit}).}
\label{Tab1}
\end{center}
\end{table*}

Figure \ref{fig:1} presents the relationship between $\gamma_s$ and $\gamma_q$ with respect to $\sqrt{s_{NN}}$. The solid symbols represent the current findings obtained from statistical fits of the measured $\mathrm{K}^+/\pi^+$ ratio to the generic (non)extensive calculations, as described by Eq. (\ref{FDn1}). On the other hand, the open symbols depict the calculations derived from the extensive hadron resonance gas (HRG) model, which is based on extensive BG statistics \cite{Tawfik:2016rrv,Tawfik:2017oyn}. The top panel of Figure 1 illustrates the variation of $\gamma_s$ with $\sqrt{s_{NN}}$ as calculated using generic (non)extensive statistics, where $\gamma_q=1$, indicating partial-chemical nonequilibrium. In the bottom panel of Fig. \ref{fig:1}, the left panel represents the dependence of $\gamma_s$ on $\sqrt{s_{NN}}$, while the right panel shows the dependence of $\gamma_q$ on $\sqrt{s_{NN}}$, indicating full-chemical nonequilibrium. The solid curves in the figure represent the proposed expression, Eq. (\ref{eq:ourFit}), based on generic (non)extensive statistics for the dependence of $\gamma_s$ and/or $\gamma_q$ on $\sqrt{s_{NN}}$. On the other hand, the double-dotted curves represent the corresponding expressions derived from the extensive HRG model, which is based on extensive BG statistics \cite{Tawfik:2016rrv,Tawfik:2017oyn}.

In top panel of Fig. \ref{fig:1}, at $\gamma_q=1$ and partial-chemical nonequilibrium, we notice that $\gamma_s$ exponentially increases, at $\sqrt{s_{NN}}\lesssim7$~GeV. The same behavior was quantitatively observed in extensive BG calculations \cite{Tawfik:2016rrv}. The rate of increase in the present calculations (generic (non)extensive statistics) is larger than that in the extensive BG statistics. At higher energies ($7\lesssim\sqrt{s_{NN}}\lesssim 20$~GeV), $\gamma_s$ decreases rapidly. In addition, the rate of decrease $\gamma_s$ in generic (non)extensive statistics is larger than that in the extensive BG statistics. At $\sqrt{s_{NN}}>20$~GeV, $\gamma_s$ becomes energy independent. At this plateau, the resulting $\gamma_s$ from both types of statistics are in good agreement with each other, $\gamma_s\simeq 0.78\pm0.01$. We conclude that with the global assumption that $\gamma_q=1$, the resulting $\gamma_s$ equally encodes the partial-chemical nonequilibrium in both types of statistics, at $\sqrt{s_{NN}}>20$~GeV. 
At lower energies, both statistical methods show significant disagreement, especially at $\sqrt{s_{NN}}\simeq7$~GeV, where the non-extensive $\gamma\_s$ is approximately $40\%$ larger than the extensive BG $\gamma\_s$. This implies that the partial-chemical nonequilibrium in the strange-quark occupation factor is about $40\%$ larger, resulting in overcast signals for strangeness enhancement, such as at the peak of the kaon-to-pion ratio. Restricting measurements, like the multiplicity of produced particles, to extensive BG statistics obscures these signals. Therefore, experimentalists can now make recommendations for future experiments using generic non-extensive statistics, especially in Run II of the STAR beam energy scan program \cite{Nonaka:2019fad,STAR:2015vvs} and future facilities, such as FAIR and NICA, to uncover essential experimental signals.

The bottom panel of Fig. \ref{fig:1} displays the non-equilibrium values of $\gamma\_s$ (left) and $\gamma\_q$ (right). In the left panel, we can observe that as $\sqrt{s\_{NN}}$ decreases below 7 GeV and $\gamma\_q$ varies, the $\gamma\_s$ calculated from generic (non)extensive statistics increases exponentially with increasing $\sqrt{s\_{NN}}$. On the other hand, the $\gamma\_s$ deduced from the extensive Hadron Resonance Gas (HRG) calculations decreases exponentially \cite{Tawfik:2017oyn}. It is evident that the rate of increase in the early stage is more rapid than the rate of decrease in the later stage. Furthermore, we find that both the maximum [generic (non)extensive statistics] and minimum (BG statistics) values are likely to be positioned at $\sqrt{s\_{NN}}\simeq 7$ GeV. Between 7 GeV and 20 GeV, the generic (non)extensive $\gamma\_s$ decreases rapidly, while the extensive BG $\gamma\_s$ increases rapidly.

For $\sqrt{s_{NN}}>20$~GeV, the energy independence of $\gamma_s$ is observed for both types of statistics. In contrast to the partial-chemical nonequilibrium scenario (top panel), the plateaus do not align. The extensive BG yields $\gamma_s\simeq 0.98\pm0.1$, indicating proximity to equilibrium, while the generic (non)extensive case results in $\gamma_s\simeq 0.81\pm0.01$, reflecting a state of significant nonequilibrium. In assessing the level of (non)equilibrium, the corresponding $\gamma_q$ shown in the bottom left panel must also be considered. Notably, at higher energies, the crest of the $\gamma_s$ curve [generic (non)extensive] surpasses the trough depth of the $\gamma_s$ curve [extensive BG] at $\sqrt{s_{NN}}\simeq 7$~GeV.

In the energy dependence analysis of $\gamma_q$ shown in the lower right panel of Fig. \ref{fig:1}, the variations of $\gamma_s$ are considered. It is evident that at $\sqrt{s_{NN}}\lesssim 7$~GeV, both generic (non)extensive $\gamma_q$ and extensive BG $\gamma_q$ (HRG model) exhibit a rapid exponential decrease with increasing energy. The rate of decrease for the generic (non)extensive $\gamma_q$ is higher than that of the extensive BG $\gamma_q$. Although the minima are closely positioned around $\sqrt{s_{NN}}\simeq 7$~GeV, their depths differ. The depth of the generic (non)extensive $\gamma_q$ surpasses that of the extensive BG $\gamma_q$. Additionally, $\gamma_q$ reaches an energy-independent state beyond $\sqrt{s_{NN}}>20$~GeV. The plateaus' heights do not coincide, with the generic (non)extensive results being lower than the extensive BG outcomes. It is noteworthy that both generic (non)extensive and extensive BG $\gamma_q$ values exceed 1.

The two lower panels demonstrate several key observations. Firstly, it is evident that the values of $\gamma_q$ and $\gamma_s$ derived from generic (non)extensive statistics are consistently lower compared to their counterparts in extensive BG statistics. Secondly, an interesting exception arises at $\sqrt{s_{NN}}\lesssim 7$~GeV, where the generic (non)extensive $\gamma_s$ attains its maximum value. Thirdly, the range of values for $\gamma_q$ and $\gamma_s$ in extensive BG statistics is narrower when compared to generic (non)extensive statistics. Lastly, at higher energies, specifically $\sqrt{s_{NN}}\gtrsim 20$~GeV, both generic (non)extensive and extensive BG $\gamma_{q,s}$ exhibit a near energy-independence, with $\gamma_q$ being greater than 1 and $\gamma_s$ being less than 1. These findings shed light on the distinct characteristics and behaviors of $\gamma_q$ and $\gamma_s$ in different statistical frameworks.

\subsection{$\mathrm{K}^+/\pi^+$ ratio}
\label{sec:kpi}

\begin{figure}[htb]
\center
\includegraphics[width=0.5\textwidth]{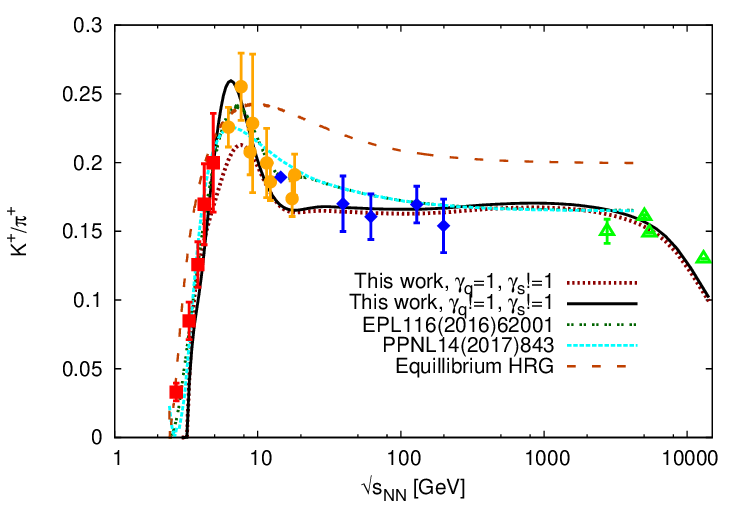}
\caption{The results from the experiments on the $\mathrm{K}^+/\pi^+$ ratio are presented as a function of $\sqrt{s_{NN}}$ in various references \cite{E866:1999ktz,NA44:2002wha,NA49:2002pzu,NA49:2004iqm,STAR:2005gfr,ALICE:2012ovd, STAR:2019vcp,ALICE:2021lsv,ALICE:2019hno}. The solid and dotted curves illustrate general (non)extensive calculations under conditions of full ($\gamma_{q}$ and $\gamma_{s}\neq1$) and partial-chemical nonequilibrium ($\gamma_s\neq1$), respectively. Meanwhile, the double-dashed and dashed curves represent the extensive BG outcomes under full \cite{Tawfik:2016rrv} and partial-chemical nonequilibrium \cite{Tawfik:2017oyn}. Lastly, the long-dashed curve showcases the HRG computations at equilibrium, where $\gamma_q=\gamma_s=1$.
}
\label{fig:2}
\end{figure}

 
Figure \ref{fig:2}  presents the $\mathrm{K}^+/\pi^+$ ratio as a function of $\sqrt{s_{NN}}$, where the symbols represent experimental results obtained at energies ranging from $5$~GeV to $5.44$~TeV \cite{E866:1999ktz,NA44:2002wha,NA49:2002pzu,NA49:2004iqm,STAR:2005gfr,ALICE:2012ovd,
STAR:2019vcp,ALICE:2021lsv,ALICE:2019hno}. The dotted curve represents the generic (non)extensive calculations under partial-chemical nonequilibrium conditions, with $\gamma_s$ varying while $\gamma_q$ remains constant at 1. On the other hand, the solid curve illustrates the results obtained under full-chemical nonequilibrium conditions, with both $\gamma_s$ and $\gamma_q$ varying. The double-dashed curve corresponds to extensive BG calculations under partial-chemical nonequilibrium \cite{Tawfik:2016rrv}, while the dashed curve represents extensive BG calculations under full-chemical nonequilibrium \cite{Tawfik:2017oyn}. The long-dashed curve demonstrates the inability of the equilibrium HRG model, where $\gamma_q=\gamma_s=1$, to reproduce both the horn structure of the $\mathrm{K}^+/\pi^+$ ratio at AGS, SPS, and low RHIC energies, as well as the measurements obtained at RHIC and LHC energies.

The HRG model examines the thermodynamic properties of an ideal gas consisting of point-like hadrons. These hadrons are characterized by the Hagedorn mass spectrum, which includes both discrete (measured) and continuous (missing) states. The analysis focuses on various higher-order thermodynamic quantities \cite{ManLo:2016pgd}. In a previous study \cite{Tawfik:2017ggu}, one of the authors (AT) concluded that the missing states do not significantly contribute to the lower-order thermodynamic quantities, particularly the first-order ones such as particle number or particle multiplicity. Consequently, the inclusion of these missing states would not enhance the thermal model's capability to reproduce the horn-structure of the $\mathrm{K}^+/\pi^+$ ratio, as detailed in Appendix \ref{sec:largemasssgma}.

The experimental results are in excellent agreement with the generic (non)extensive calculations across the entire energy spectrum. The horn-like structure of the kaon-to-pion ratio is accurately reproduced. In comparison to other statistical approaches, the generic (non)extensive method is highly effective in describing the rapid increase at $\sqrt{s_{NN}}\lesssim 7$~GeV and the subsequent fast decrease at $7\lesssim\sqrt{s_{NN}}\lesssim 20$~GeV. At $\sqrt{s_{NN}}\gtrsim 20$~GeV, a detailed plateau structure is predicted. This estimation is crucial due to the significant energy gap between the top RHIC and LHC energies. Additionally, it is observed that both partial- and full-chemical nonequilibrium outcomes exhibit significant variations, particularly at $\sqrt{s_{NN}}\lesssim 7\pm1$~GeV. The experimental results are in excellent agreement with the generic (non)extensive calculations across the entire energy spectrum. The horn-like structure of the kaon-to-pion ratio is accurately reproduced. In comparison to other statistical approaches, the generic (non)extensive method is highly effective in describing the rapid increase at $\sqrt{s_{NN}}\lesssim 7$~GeV and the subsequent fast decrease at $7\lesssim\sqrt{s_{NN}}\lesssim 20$~GeV. At $\sqrt{s_{NN}}\gtrsim 20$~GeV, a detailed plateau structure is predicted. This estimation is crucial due to the significant energy gap between the top RHIC and LHC energies. Additionally, it is observed that both partial- and full-chemical nonequilibrium outcomes exhibit significant variations, particularly at $\sqrt{s_{NN}}\lesssim 7\pm1$~GeV. 

The full-chemical nonequilibrium hypothesis suggests that there will be a further increase in the $\mathrm{K}^+/\pi^+$ ratio, especially when the center-of-mass energy $\sqrt{s_{NN}}$ is between 3 and 10 GeV. To refine the conclusion regarding whether partial- or full-chemical nonequilibrium accurately reproduces the nonmonotonic horn-structure of the $\mathrm{K}^+/\pi^+$ ratio at AGS, SPS, and low RHIC energies, it is highly recommended to conduct a comprehensive beam energy scan using the STAR and CBM experiments. This scan will provide valuable insights and contribute to a better understanding of the phenomenon.

\section{Conclusion}
\label{sec:Diss_concl}

This research paper presents a novel concept of generic (non-extensive) statistics. The main objective is not only to propose an approximate solution to a long-standing problem, but also to address the shortcomings of thermal models in accurately reproducing the nonmonotonic horn-like structure of the $\mathrm{K}^+/\pi^+$ ratio. The assumption that particle production can be analyzed using extensive additive equilibrium statistics contradicts the inherent characteristics of high-energy particle production. Although this assumption could be accepted in the absence of any alternative, the non-extensive Tsallis approach also proves to be inadequate. In light of these limitations, this paper introduces a first-principle solution that is based on a proper statistical approach. This approach ensures that the extensivity and nonextensivity of the statistics align well with the statistical nature of particle production. 

Several factors have been introduced to enhance the effectiveness of thermal approaches in characterizing particle ratios. These include modifying the single-particle distribution function \cite{Tawfik:2010uh}, considering nonequilibrium light- and strange-quark occupation in the phase-space volume (i.e., $\gamma_{q,s}\neq 1$) \cite{Tawfik:2005gk,Tawfik:2016rrv,Tawfik:2017oyn}, and incorporating an excluded volume in the pointlike ideal gas of hadron resonances \cite{Tawfik:2014eba,Tawfik:2013eua}. These modifications aim to improve the capability of thermal approaches, which are based on either extensive additive BG statistics \cite{Gazdzicki:1998vd,Cleymans:2005zx,Andronic:2005yp,Tawfik:2005gk,Tomasik:2006qs,Chatterjee:2009km,Tawfik:2010uh,Nayak:2010uq} or Tsallis-type nonextensive statistics \cite{Tawfik:2019alr}. However, these attempts have not fully succeeded in accurately characterizing particle ratios, including the nonmonotonic horn-like structure observed in the $\mathrm{K}^+/\pi^+$ ratio. This structure likely indicates the onset of non-extensive particle production and the influence of non-equilibrium phase transition \cite{Tawfik:2018sji}. 

The current manuscript posits that the inability of thermal models to replicate a horn-like structure (the long-dashed curve in Fig. \ref{fig:2}) is attributed to inappropriate statistics, such as extensive additive BG or Tsallis-type nonextensive statistics, or their variations. Introducing nonequilibrium components like a modified single-particle distribution function, nonequilibrium occupation factors $\gamma_{q,s}$, or finite volume of the constituents of the ideal gas to extensive additive BG statistics is deemed conceptually unproductive. Nevertheless, they are suitable for non-extensive statistics. In the context of (non)extensivity, Tsallis-type nonextensivity is confined to a line, where $c=0$ and $d=q$. Conversely, general (non)extensive statistics encompass the entire space of the two equivalence classes $(c,d)$. Through this statistical framework, the ensemble establishes its statistical extensivity and non-extensivity. Both extensive BG and Tsallis-type non-extensive statistics were encompassed, characterized by $(1,1)$ and $(0,q)$, respectively.

Firstly, the determination of partial- and full-chemical nonequilibrium, denoted as $\gamma_q$ or/and $\gamma_s$, has been conducted at various energies. Secondly, in a similar manner, the estimation of the energy-dependence of $(c,d)$ has been performed. When calculating the $\mathrm{K}^+/\pi^+$ ratio, the occupation factors $\gamma_{q,s}$, the equivalence classes $(c,d)$, and the freezeout parameters ($T$, $\mu_B$) are no longer considered as fitting variables. Instead, they are treated as inputs at different energies. It has been observed that the $\mathrm{K}^+/\pi^+$ ratio calculated using generic (non)extensive statistics remarkably reproduces the experimental results, including the horn-like structure. The nonmonotonic energy-dependence of $\gamma_{q,s}$ and $(c,d)$ is crucial in achieving this excellent reproduction. In conclusion, the accurate reproduction of the $\mathrm{K}^+/\pi^+$ ratio necessitates the consideration of the aforementioned factors. 
\begin{enumerate}
\item Lambert-$W_0$ exponentially generated distribution rather than BG or Tsallis  and 
\item partial- or full-chemical nonequilibrium occupation factors. 
\end{enumerate}
Within this framework, we can confirm that various ratios, including proton-to-pion, are faithfully reproduced, as detailed in Appendix \ref{sec:cdcountors}.

While the measured $\mathrm{K}^+/\pi^+$ ratio is excellently reproduced by both partial- and full-chemical nonequilibrium in generic (non)extensive statistics across all energies, it is worth noting that the full-chemical nonequilibrium appears to be particularly significant within the energy range of $3\lesssim\sqrt{s_{NN}}\lesssim 10$~GeV. In order to further investigate and refine the conclusion regarding the essentiality of partial- or full-chemical nonequilibrium in explaining the nonmonotonic horn-structure of the $\mathrm{K}^+/\pi^+$ ratio, it is highly recommended to conduct a detailed beam energy scan using the STAR and CBM experiments, among others.

\appendix

\section{Probability distribution, entropy, exponential and logarithm functions in Boltzmann, Tsallis and generic (non)extensive statistics}
\label{sec:expf}

In order to establish a trustworthy comparison between Boltzmann, Tsallis, and generic (non)extensive statistics, it is imperative to revisit the corresponding probability distributions \cite{Tawfik:2023xgz}
\bea
\mathtt{Extensive\; BG:} && p_i(x) = \frac{\exp(-x)}{{\cal Z}(x)}= \frac{1}{{\cal Z}(x)}  \sum_{n=0}^{\infty}\frac{x^n}{n!},  \\
\mathtt{Nonextensive\; Tsallis:} && p_i^{(q)}(x) = \frac{\exp^{(q)}(-x)}{{\cal Z}^{(q)}(x)}= \frac{1}{{\cal Z}^{(q)}(x)} [1-x+\frac{q}{2}x^2 - \frac{q^2}{3!} x^3 + \cdots] \nn \\
&&\hspace*{12mm}=  \frac{1}{{\cal Z}^{(q)}(x)}  \left[1+\sum_{n=1}^{\infty}\frac{x^n}{n!}Q_{n-1}(q)\right] = \frac{1}{{\cal Z}^{(q)}(x)} \left[1+(1-q)x\right]^{1/(1-q)},  \hspace*{8mm}\\
\mathtt{(Non)extensive:} && p_i^{(c,d,r)}(x) = \frac{\exp^{(c,d,r)}(-x)}{{\cal Z}^{(c,d,r)}(x)} \nn \\
&&\hspace*{17mm} = \frac{1}{{\cal Z}^{(c,d,r)}(x)}
\exp\left[\frac{-d}{1-c}\left\{W_k\left[B\left(1-\frac{x}{r}\right)^{1/d}\right]\right\}-W_k[B]\right], 
\eea
where ${\cal Z}(x)$ is the corresponding canonical partition function, $Q_{n-1}(q):=\Pi_{i=0}^n [iq-(i-1)]$ \cite{Borges:1998xr}, and the asymptotic expansion of $0$-th branch Lambert-$W_k$ function is given as 
\bea
W_{k=0}(x) = \sum_{n=1}^{\infty} (-1)^{n-1}\frac{n^{n-1}}{n!}\; x^n. \nn
\eea

The probability distribution function can be derived by integrating the probability density function, which quantifies the probability of a particular event occurring within a defined interval. Furthermore, the probability distribution function can be obtained by maximizing the corresponding entropy subject to appropriate constraints \cite{Tawfik:2023xgz}
\bea
\mathtt{Extensive\; BG:} && s[p] = - \kappa \sum_{i=1}^{\Omega} p_i\; \ln\; p_i,  \\
\mathtt{Nonextensive\; Tsallis:} && s^{(q)}[p] = \frac{\kappa}{1-q} \sum_{i=1}^{\Omega} \left(p_i^{(q)}-p_i\right),\nn \\
\mathtt{(Non)extensive:} && s^{(c,d,r)}[p] = \kappa \sum_{i=1}^{\Omega} \left[{\cal A}\, \Gamma(d+1, 1 - c \log p_i) - {\cal B}\, p_i\right], \label{eq:NewExtns1}
\eea
where $\Omega$ is the number of micro-states or processes. $\kappa$, ${\cal A}$, and ${\cal B}$ are arbitrary parameters. 
The incomplete gamma-function is given as \cite{Pereira2012AnalyticalSO}
\bea
\Gamma (a, x)=\int_{x}^{\infty}\, dt\, t^{a-1} \exp (-t) &=& \Gamma(a)- \frac{x^a}{a\, \exp(x)} \sum^{\infty}_{i=0} \frac{x^i(a)!}{(a-i)!}.
\eea

The distinctions among Boltzmann, Tsallis, and generic (non)extensive statistical statistics can be distinguished by the mathematical interpretation of the corresponding exponential and logarithm functions. Specifically, for Boltzmann, the functions are represented as $\exp(x)$ and $\ln(x)$, for Tsallis as $\exp^{(q)}(x)$ and $\ln^{(q)}(x)$, and for generic (non)extensive functions as $\exp^{(c,d,r)}(x)$ and $\ln^{(c,d,r)}(x)$. An illustration of this can be seen in the Tsallis exponential and logarithm functions as mentioned in \cite{Umarov2008OnAQ}
\bea
\exp^{(q)}(-x) &=& 1+\sum_{n=1}^{\infty} (-1)^n \frac{q^{n-1}}{n!}x^n= 1 - x+\frac{q}{2!}x^2 - \frac{q^2}{3!} x^3 + \cdots, \label{wq:TslsExp}\nn \\
\ln^{(q)}(x) &=& \sum_{n=1}^{\infty}(-1)^{n+1}\frac{q^{n-1}}{n}(x-1)^n=(x-1) - \frac{q}{2} (x-1)^2 + \frac{q^2}{3}(x-1)^3 - \cdots. \label{wq:TslsLn}
\eea
For generic (non)exrensive exponential and logarithm functions, generaltization of BG  $\exp(x)$ and $\ln(x)$ gives \cite{Tawfik:2016jol}
\bea
\exp^{(c,d,r)}(x) &=& \exp\left\{\frac{-d}{1-c}\left[\mathtt{W}_k\left(B\left(1-x/r\right)^{1/d}\right) - \mathtt{W}_k(B)\right]\right\}, \hspace*{3mm} \label{eq:ps1} \\
\ln^{(c,d,r)}(x) &=& r\, x^{c-1} \left[1-\frac{1-(1-c) r}{r\, d} \log(x)\right]^d. \label{eq:genLog}
\eea

It is now appropriate to provide an analysis of the interpretations of the Tsallis nonextensitivity parameter $q$. Initially, the assumption that $q>1$ indicates fluctuations in $1/\lambda$ of the Levy exponential distribution, $\exp(-x/\lambda)$, which may seem superficial \cite{Wilk:1999dr}. However, the nature of these fluctuations has not yet been determined. Additionally, these fluctuations should also result in a significant alteration in the phase space itself \cite{Tawfik:2010uh}. Furthermore, we evaluate the relationship between the deviation of $q$ from its extensive value (unity) from above and the occurrence of large temperature fluctuations. 

Moreover, the assertion that there are fundamental distinctions between BG- and Tsallis-statistics\footnote{There is no fundamental difference between BG- and Tsallis-statistics, unless the earlier is an extensive and the latter is a nonextensive statistical framework!} necessitates an arbitrary prerequisite. This prerequisite suggests that the thermodynamic temperature, derived from BG statistics, differs from the temperature determined in Tsallis statistics, $T_{Tsallis}=T_{BG}+(q-1)K$, where the parameter $K$ should be modeled as a function of the energy transfer between the source and the surrounding \cite{Deppman:2012qt}.

This particular interpretation, grounded in phenomenology, is directly connected to the flawed utilization of Tsallis-algebra, which involves the substitution of $\exp(x)$ with $\exp_q(x)$ as shown in Eq. (\ref{wq:TslsExp}), and $\ln(x)$ with $\ln_q(x)$ as shown in Eq. (\ref{wq:TslsLn}). Notably, there was a misinterpretation in replicating the transverse momentum distributions, leading to an inaccurate determination of the resulting temperature as reported in \cite{Bialas:2015pla}.

The current manuscript posits that Boltzmann and Tsallis statistics fall under the umbrella of generic (non)extensive statistics. The Tsallis nonextensive parameter $q$ is a part of both Boltzmann statistics when $q=1$ and generic (non)extensive statistics when $c=0$ and $d=q$. The equivalence classes $(c,d)$ are inherent characteristics of the statistical space as a whole, with $(c,d)=(1,1)$ representing Boltzmann statistics and $(c,d)=(0,q)$ corresponding to the Tsallis type. Thermodynamic quantities, whether intensive or extensive, such as temperature, entropy, pressure, and energy density, should remain consistent throughout the statistical space, ensuring that they do not vary when transitioning from Boltzmann statistics to Tsallis statistics or any other form of nonextensive statistics. The interpretations of $q$ and $(c,d)$ go beyond mere random fluctuations, serving as nonextensive statistical descriptors that are linked to the properties of the specific statistical ensemble.

\section{Inclusion of very high-mass resonances and missing states in the hadron resonance gas model}
\label{sec:largemasssgma}

\begin{figure}[htb]
\center
\includegraphics[width=0.5\textwidth]{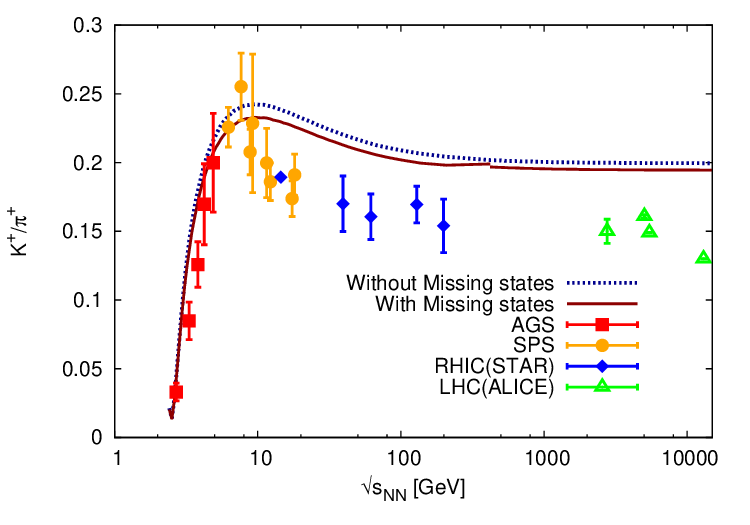}
\caption{The energy dependence of the relative production ratio $\mathrm{K}^+/\pi^+$ was investigated through measurements at various energy levels. The $\mathrm{K}^+/\pi^+$ ratio was estimated using the hadron resonance gas model, incorporating extensive additive equilibrium BG statistics. The degrees of freedom (dof) were determined based on the mass-truncated PDG compilation of hadrons and resonances. The estimated ratios were illustrated with a dotted curve.
}
\label{fig:0}
\end{figure}

Two key additions were made by the authors in reference \cite{Andronic:2008gu} to the extensive additive equilibrium hadron resonance gas model. The first addition pertains to high-mass resonances with masses exceeding 2 GeV. The second addition involves the $\sigma$ meson or $f_0(600)$. Furthermore, the authors took into consideration the various decay channels in their analysis.

The $\sigma$ meson is characterized by a mass that varies with energy and decay constants that differ. Its primary decay channel results in the production of a pair of pions. This observation may have served as the motivation for reference \cite{Andronic:2008gu}. It has been estimated that the $\sigma$ meson contributes to an approximately $3.5\%$ increase in pion yields. Furthermore, the inclusion of high-mass resonances also leads to an increase in pion yields, this time by around $13\%$. By prioritizing the enhancement of pion yields over kaon yields, it is possible to reproduce the nonmonotonic $\mathrm{K}^+/\pi^+$ ratio. However, it is important to note that the standard of evidence requires a similar enhancement of kaon yields. Alternatively, the ensemble of hadron resonances remains unbalanced and biased.

Figure \ref{fig:0} illustrates an unbiased balanced representation of the $K^+/\pi^+$ ratio, incorporating all hadrons and resonances documented in the recent PDG, including the $\sigma$ meson and its higher resonances, as well as 56 missing states. The analysis accounts for most, if not all, decay channels, ensuring that the final count of any particle considers the particle itself and all other decay channels that produce it, weighted by the corresponding decay width. The masses of the resonances can reach up to 12 GeV. Our findings suggest that the incorporation of very heavy-mass resonances marginally boosts the respective particle yields. However, this enhancement diminishes in the relative production ratio. The $K^+/\pi^+$ ratio is not accurately replicated with or without the inclusion of missing states.

\begin{figure}[htb]
\center
\includegraphics[width=0.45\textwidth]{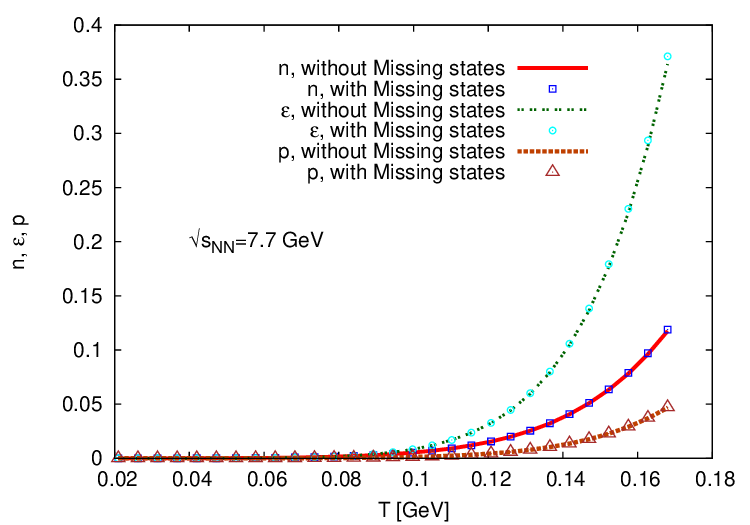}
\includegraphics[width=0.45\textwidth]{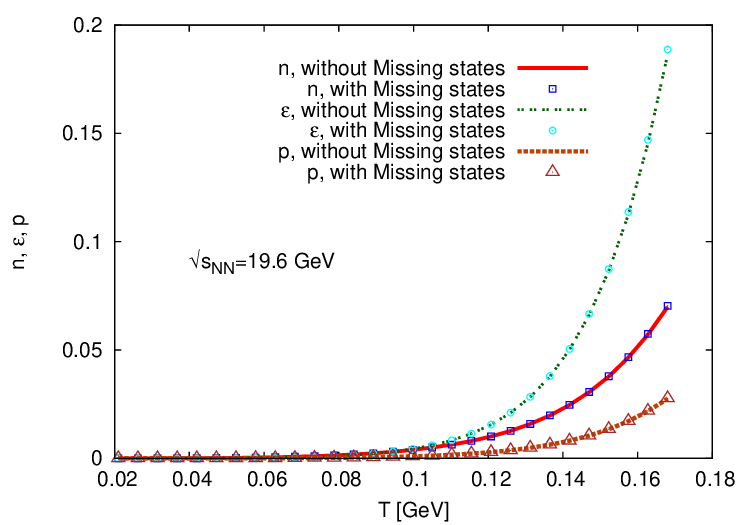}\\
\includegraphics[width=0.45\textwidth]{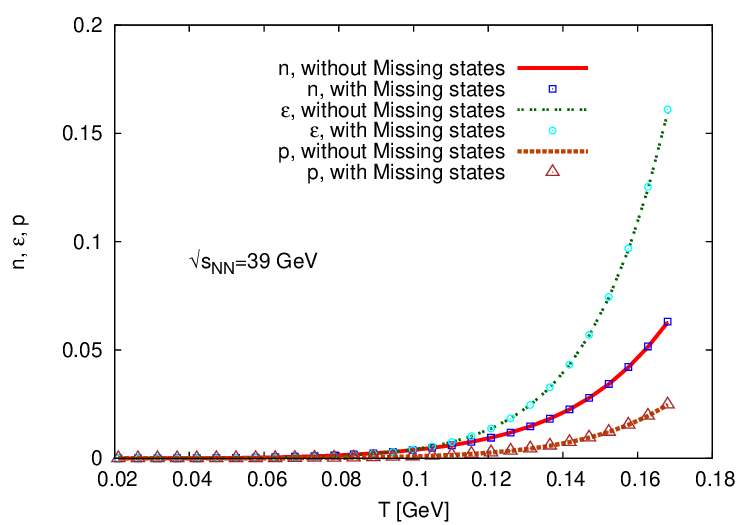}
\includegraphics[width=0.45\textwidth]{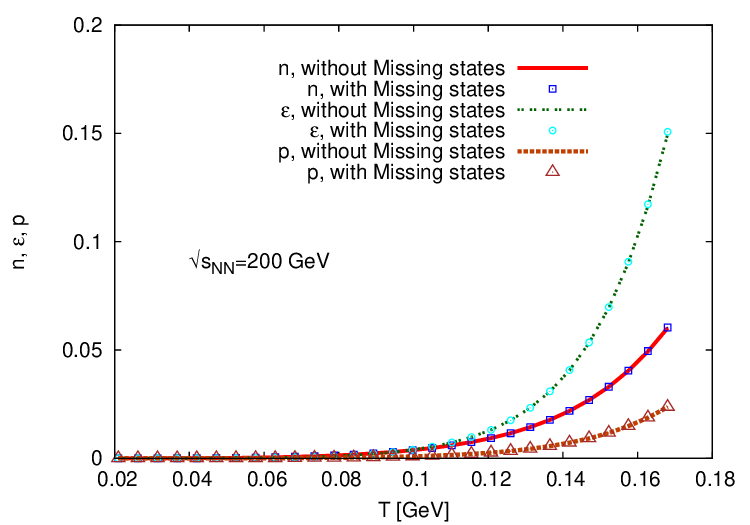}
\caption{In the extensive additive equilibrium hadron resonance gas model, the temperature dependence of number density, energy density, and pressure was computed for energies of $7.7$, $19.6$, $39$, and $200~$GeV. The outcomes are illustrated through curves, which exhibit the effects of truncating the hadron mass spectrum at $<2~$GeV. The symbols employed in the figures represent the results that encompass $56$ missing states and accommodate hadron masses up to $12~$GeV.
}
\label{Fig:nnn}
\end{figure}

The Fig. \ref{Fig:nnn} illustrates the contributions of the missing states to various thermodynamic quantities \cite{ManLo:2016pgd}. Specifically, at energy levels of $7.7$, $19.6$, $39$, and $200~$GeV, the number density, energy density, and pressure obtained from the extensive additive equilibrium hadron resonance gas model are compared. The comparison is made between calculations that include (represented by symbols) and exclude (represented by curves) missing states, as well as between calculations that include (symbols) and exclude (curves) very-heavy mass resonances. The analysis reveals that the impact of including the missing states and very-heavy mass resonances on particle yields is minimal. These contributions seem to be balanced in the relative particle production ratios, as shown in Fig. \ref{fig:0}. It is worth noting that the authors of ref. \cite{ManLo:2016pgd} have also reached the same conclusion, stating that the missing resonances in the strange baryon sector are responsible for the susceptibilities of conserved charges. In other words, the missing states primarily affect higher-order moments of conserved charges, such as susceptibilities and cumulants, while their influence on lower-order moments, such as the number density, is negligible. Furthermore, even this slight contribution to lower-order moments is overshadowed in the context of particle ratios.

\section{Contour plots of the quality of fits for $\gamma_{q,s}$} 
\label{sec:cdcountors}

\begin{figure}[htb]
\center
\includegraphics[width=0.45\textwidth]{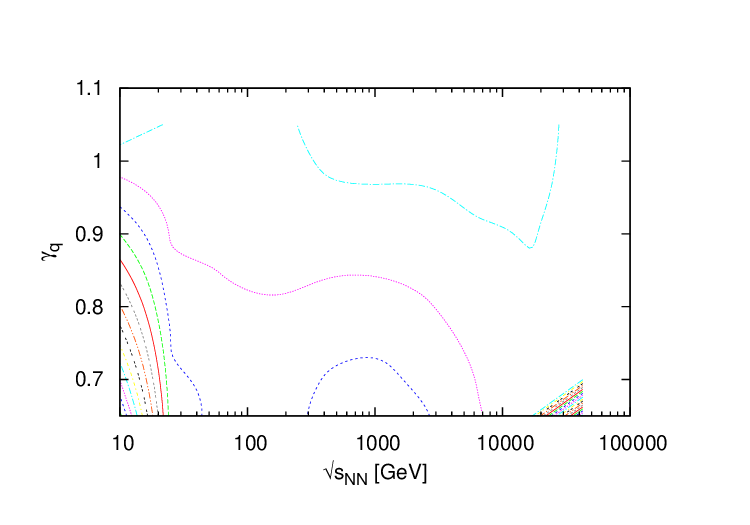}
\includegraphics[width=0.45\textwidth]{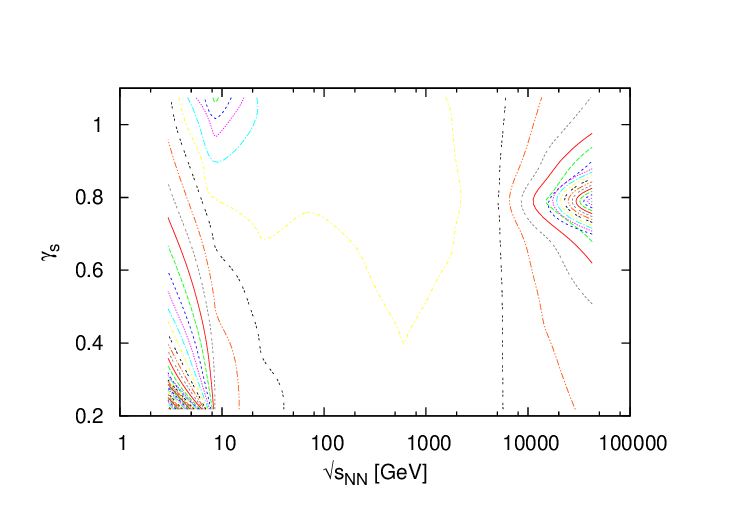}\\
\includegraphics[width=0.5\textwidth]{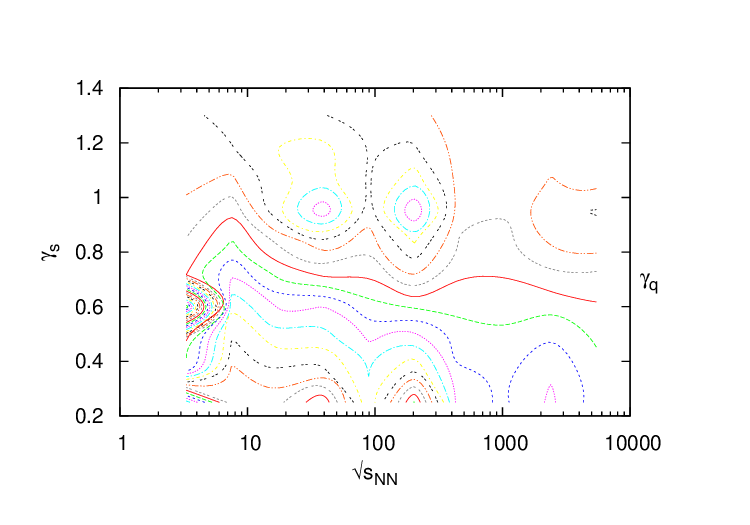}
\caption{The contour plot displayed in the top left represents the quality of fit for $\gamma_q$ at various collision energies, with $\gamma_s$ fixed at 1. Similarly, the top right contour plot illustrates the quality of fit for $\gamma_s$ while keeping $\gamma_q$ constant at 1. Moving on to the bottom panel, it showcases a contour plot for nonequilibrium $\gamma_q$ and $\gamma_s$ at different collision energies.
}
\label{Fig:cdcontors}
\end{figure}

A decrease in $\gamma_s$ was considered as an indication of the beginning of color deconfinement. The non-equilibrium quark occupation factors, $\gamma_{q,s}\neq1$, are believed to determine the quantities of light- and strange-quarks that can be contained in the phase-space volume, thus affecting the production rates of light and strange particle yields. Various studies have employed extensive additive equilibrium BG statistics for these investigations.

When $\gamma_{q}$ is determined based on the statistical fit of one particle ratio, the reproduction of other particle ratios is not well achieved. Despite the possibility of simultaneously reproducing various other particle ratios for nonequilibrium $\gamma_{q,s}$, the resulting values of $\gamma_{q,s}$ exhibit scattering, as presented in Table 1 of reference \cite{Tawfik:2005gk}. Incorporating values of $\gamma_{q,s}$ that are significantly greater than unity poses challenges within the context of a nonequilibrium approach.

The contour plots in Fig. \ref{Fig:cdcontors} depict the quality of fits for three parameters: $\gamma_q$ (top left panel), $\gamma_s$ (top right panel), and $\gamma_{q,s}$ (bottom panel). The parameterization of $\gamma_{q,s}$ is given by Eq. (\ref{eq:ourFit}). Upon comparing these estimations with the data presented in Table 1 of reference \cite{Tawfik:2005gk}, it can be inferred that the current estimations align well with an nonequilibrium approach. 

\begin{figure}[htb]
\center
\includegraphics[width=0.45\textwidth]{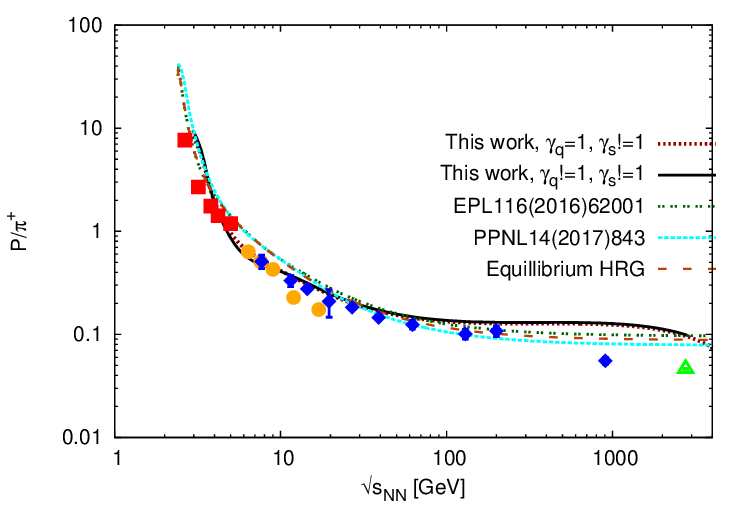}
\includegraphics[width=0.45\textwidth]{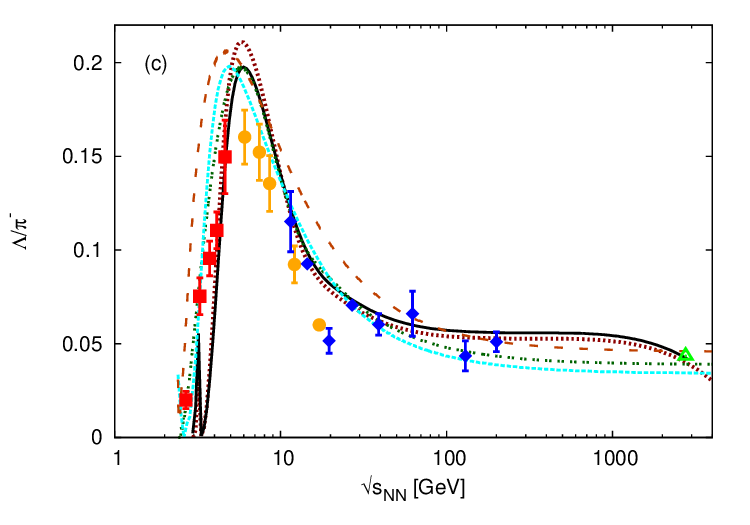}
\caption{The energy dependence of $\mathrm{P}/\pi^+$ is shown in the left panel, while the right panel illustrates the energy dependence of $\Lambda/\pi^-$. The equivalence classes are determined by Eqs. (\ref{eq:c})-(\ref{eq:d}), and $\gamma_{q,s}$ is defined by Eq. (\ref{eq:ourFit}), mirroring the approach taken for the analysis of $\mathrm{K}^+/\pi^+$.
}
\label{fig:Ppip}
\end{figure}

By utilizing the same sets of equivalence classes given by Eqs. (\ref{eq:c})-(\ref{eq:d}) and the nonequilibrium quark-occupation factors $\gamma_{q,s}$ given by Eq. (\ref{eq:ourFit}), not only is the $\mathrm{K}^+/\pi^+$ ratio accurately reproduced, but other particle ratios such as $\mathrm{P}/\pi^+$ and $\Lambda/\pi^-$ are also well reproduced. This can be observed in Fig. \ref{fig:Ppip}. It is important to emphasize that the current ensemble of experimental results available at energies above $4~$TeV is insufficient to determine the equivalent classes.

\bibliographystyle{aip}
\bibliography{mybibfile}

\end{document}